\documentclass[preprint2]{aastex}
\usepackage{graphicx}
\usepackage{amssymb}
\usepackage{natbib}

\begin{document}
\title{PReBeaM for \textsc{Planck}: A Polarized Regularized Beam Deconvolution Map-Making Method}

\author{Charmaine Armitage-Caplan\altaffilmark{1} and Benjamin D.~Wandelt
\altaffilmark{1,2,3}}

\altaffiltext{1}{Department of Physics, UIUC, 1110 W Green Street, Urbana, IL 61801}
\altaffiltext{2}{Department of Astronomy, UIUC, 1002 W Green
Street, Urbana, IL 61801}
\altaffiltext{3}{Center for Advanced Studies, UIUC, 912 W Illinois Street, Urbana,
  IL 61801}

\begin{abstract}
We describe a maximum likelihood regularized beam deconvolution map-making algorithm for 
data from high resolution, polarization sensitive instruments, such as the \textsc{Planck}
 data set.
The resulting algorithm, which we call PReBeaM, is pixel-free and solves for the map directly 
in spherical harmonic space, avoiding pixelization artifacts.
While Fourier methods like ours are expected to work best when applied to smooth, large-scale 
asymmetric beam systematics (such as far-side lobe effects) we show that our $m$-truncated 
spherical harmonic representation of the beam results in negligible reconstruction error 
-- even for $m$ as small as 4 for a polarized elliptically asymmetric beam.
We describe a hybrid OpenMP/MPI parallelization scheme which allows us to store and manipulate 
the time-ordered data from instruments with arbitrary scanning strategy.
Finally, we apply our technique to noisy data and show that it succeeds in removing visible 
power spectrum artifacts without generating excess noise on small scales.
\end{abstract}

\section{Introduction}

One of the most exciting prospects for the upcoming \textsc{Planck} satellite
is its capability to measure the polarization anisotropies of the CMB over 
the entire sky in nine frequency channels.  The potential 
rewards from these measurements are many and include tighter constraints on cosmological
parameters, determination of the reionization history of the universe, and detection
of signatures left by primordial gravitational waves generated during inflation \citep{bluebook}.

Measurement of the CMB polarization signal presents a great experimental 
challenge as it is an order of magnitude smaller than the temperature signal 
and is especially susceptible to distortions due to optical systematics and
foreground contaminants.  
Indeed, if left untreated, leakage from the much stronger temperature signal will
contaminate the polarization maps.  Maps and spectra will also suffer from leakage from
E-mode polarization to B-mode polarization, jeopardizing the potential detection
of inflationary B-modes.  At the resolution and sensitivity of the next generation 
of experiments, including the Planck mission, studies of primordial non-Gaussianity 
may also be sensitive to beam-induced systematics.
In this paper, we present a novel technique for both assessing and removing
systematic effects due to beams in temperature and polarization maps.

The \textsc{Planck} satellite is designed 
to extract essentially all of the information in the primordial temperature anisotropies and 
to measure the polarization anisotropies to high accuracy for $2\lesssim\ell\lesssim 2500$.  
This will be achieved 
by measuring the full-sky signal to an angular resolution of 5', to a sensitivity of 
$\Delta$T/T $\sim 2$x$10^{-6}$, and over a frequency range of 30-857 GHz \citep{bluebook}.
The scientific performance of \textsc{Planck} depends, in part, on the behavior of systematic
effects which may distort the signal.  

A primary objective of \textsc{Planck} is to produce
all-sky CMB maps at each frequency.  The process by which the satellite's time-ordered data (TOD)
is wrapped back on to the sphere to create an image is known as map-making.  The
map-making process becomes difficult due to a number of challenges: distortions in the beam, 
foreground contamination through far-side lobes, size of the data, and correlated noise effects.
It is of critical importance to fully
characterize the beam, and use this information during map-making to
deconvolve beam effects.  We have previously described a powerful map-making 
algorithm which implements the beam deconvolution technique for temperature 
measurements \citep{AW04}.  In this paper, we will extend that description 
to include {\it polarization} measurements.  We refer to this new technique
as PReBeaM: Polarized Regularized Beam deconvolution Map-making.
While we focus on reconstructing the map with a uniform effective beam and 
realize corrections to the power spectrum as a consequence, other work 
by \citet{Souradeep06} and \citet{Mitra07} has 
focused on deriving corrections to the power spectrum due to asymmetric 
(non-circular) beam effects.

Within the \textsc{Planck} collaboration, the CTP working group has developed
five map-making methods and
compared their results using the simulated 30 GHz data in what is known as the Trieste paper
\citep{Trieste}.
The Trieste paper 
assessed the impact of beam asymmetries on the \textsc{Planck} spectra without attempting
to treat the problem of beam asymmetry at the map-making level (an angular power spectrum 
correction method was developed based on simplifying assumptions).
In addition to PReBeaM, another deconvolution map-making
technique for \textsc{Planck} has been established by \citet{Harrison08}. 
Both methods allow for arbitrary beam shapes and in both cases the
asymmetry of the beam is parametrized by an asymmetry parameter  $m_{max}$ which can vary
between 0 and $\ell_{max}$.  Our method scales 
computationally as a function
of $m_{max}$; this is advantageous when large gains in accuracy can be achieved with small
increases in $m_{max}$.  In contrast, the Harrison method incurs a fixed  
computational expense
for arbitrarily large $m_{max}$. 
The Harrison method takes advantage of the Planck scanning strategy
to condense the full TOD into phase-binned rings, thereby
achieving a significant reduction in processing time.  

A complete characterization of the beam includes both the main beam and the
far-side lobes.  Sidelobes are located as far away as $90^{\circ}$ from the main focal plane
beam, and therefore require a large $m_{max}$ 
parameter for a complete harmonic description.  In \citet{AW04} we demonstrated 
the full potential of our
method using far-side lobes and maps with foreground signals.  Here, we show the 
usefulness of PReBeaM
for deconvolving main-beam distortions.  In fact, we find that it makes 
sense to use PReBeaM
for main beam effects since only a small $m_{max}$ parameter is needed to capture the azimuthal
structure of the main beam.  In this way, we profit from the computational advantage 
of our method 
in the case of small $m_{max}$, allowing for the unified treatment of main beam and side lobe 
effects.

In \S\ref{sec:deconv} we describe the deconvolution map-making algorithm for PReBeaM. The simulated data and beams are detailed in \S\ref{sec:sims}.
We present results in \S\ref{sec:results} showing the effectiveness of PReBeaM
in removing systematic effects due to beam asymmetry and we discuss computational considerations.  
We finish with our conclusions from this study in \S\ref{sec:conclusion}.

\section{PReBeaM Method}
\label{sec:deconv}
First we review the standard set-up to the map-making problem for a solution
of the least-squares (or maximum-likelihood) type.

The TOD generated by a detector is effectively a
convolution of the true CMB sky with a beam function.  If we consider the sky as
a pixelized vector, it will have length $n_{pixel}\times n_{pol}$ where $n_{pol}=3$
for the I (total intensity), Q, and U Stokes components.  
The $n_{TOD}$-length TOD vector {\bf d} is the result
of a matrix multiplication of the observation matrix {\bf A} with the sky {\bf s}
\begin{equation}
\mathbf{A} \mathbf{s} = \mathbf{d}.
\end{equation}
In our implementation of the maximum-likelihood solution, we refer to {\bf A} as the convolution operator.
{\bf A} encodes information about both the scanning strategy and
the optics of the scanning instrument.
The least-squares estimate of the true sky, ${\bf \hat{s}}$, is given by the
normal equation
\begin{equation}
\label{eq:normal}
\mathbf{A}^{\mathrm{T}} \mathbf{A} \mathbf{\hat{s}} = \mathbf{A}^{\mathrm{T}}
\mathbf{d}
\end{equation}
where ${\mathbf A^{\mathrm{T}}}$ is the transpose convolution operator.
Equation (\ref{eq:normal}) is exact if the noise is stationary and uncorrelated
in the time-ordered domain.  The generalization to non-white noise is as follows
\begin{equation}
\mathbf{A}^{\mathrm{T}}\mathbf{N}^{-1} \mathbf{A} \mathbf{\hat{s}} = 
\mathbf{A}^{\mathrm{T}} \mathbf{N}^{-1}\mathbf{d}
\end{equation}
where {\bf N} is a noise covariance matrix.  In this work we consider CMB only and CMB plus white noise.

We modify the normal equation by introducing a regularization technique 
in order cope with the ill-conditioned nature
of the coefficient matrix {\bf A}$^{\mathrm T}${\bf A}.  We split off the ill-conditioned
part of {\bf A} by factoring it into two parts: {\bf A} = {\bf BG}.  
The factor {\bf G} is what we refer to as the {\it regularizer} in PReBeaM.
In general, the regularizer can be any target beam; a natural 
choice would be the angle-averaged detector beam.
In our study, we choose {\bf G} to be a
Gaussian smoothing matrix, defined in harmonic space as
\begin{eqnarray}
\label{eq:regular}
G_{\ell}^I & = & \exp\left(\frac{-\sigma^2\ell(\ell+1)}{2}\right)\nonumber\\
G_{\ell}^{G,C} & = & \exp\left(\frac{-\sigma^2(\ell(\ell+1)-4)}{2}\right)
\end{eqnarray}
where $\sigma$ = FWHM/$\sqrt{8\ln2}$.  
The superscripts $G$ and $C$ refer to the gradient and curl
components in the typical linear polarization decomposition.
Our modified normal 
equation becomes
\begin{equation}
\mathbf{B}^{\mathrm{T}} \mathbf{B} \mathbf{\hat{x}} = \mathbf{B}^{\mathrm{T}}
\mathbf{d}
\end{equation}
where we are solving for {\bf x} = {\bf G}$\mathbf{\hat{s}}$.  In this way, we do not 
attempt to reconstruct the sky at a higher resolution than that of the instrument.

In a standard pixel-based solution of equation (\ref{eq:normal}), in which one assumes
that the observing beam is spherically symmetric, ${\mathbf A}$ is a sparsely-filled
pointing matrix.  For polarization measurements, each row of ${\mathbf A}$ contains
only three non-zero elements.  The deconvolution map-making approach does not 
assume spherically symmetric beams, instead allowing for arbitrary beam shapes.  
We achieve this added complexity primarily by solving the normal 
equation in spherical harmonic space in order to make use of fast and exact algorithms 
for the convolution and transpose convolution of two arbitrary functions on the sphere \citep{WG01,C00}.
These algorithms are described in abbreviated form below.  A secondary advantage of operating
entirely in harmonic space is that artifacts due to pixelization (such as uneven sampling of the pixel) 
are completely avoided.

\subsection{Fast all-sky convolution for polariametry measurements}
For a full  presentation of the formalism for convolution of an 
instrument beam with a sky signal, the reader is referred to \citet{C00}.

In compact spherical harmonic basis, equation (\ref{eq:normal}) is written as
\begin{equation}
\label{eq:norm_harm}
A^T_{L'M'mm'm''}A_{mm'm''LM} s_{LM} = A^T_{L'M'mm'm''}T_{mm'm''}
\end{equation}
where $s_{LM}$ is the spherical harmonic representation of the sky and $T_{mm'm''}$
is defined as the result of a convolution of a band-limited function 
b with the sky s.  
The \textsc{Planck} Level-S software \citep{R06} nomenclature refers to $T_{mm'm''}$ as a 
{\it ring set}.  
This is written in harmonic space as
\begin{eqnarray}
\label{eq:conv_harm}
T_{mm'm''} = \sum_{\ell} (\frac{1}{2}s_{\ell m}^I b_{\ell M'}^{I\ast} + 
s_{\ell m}^G b_{\ell M'}^{G\ast} \nonumber \\+ s_{\ell m}^C b_{\ell M'}^{C\ast})
d_{mM}^{\ell}(\theta_E) d_{MM'}^{\ell}(\theta)
\end{eqnarray}
where $(\theta_E,\theta)$ are fixed parameters which define the scanning geometry.

In equation (\ref{eq:conv_harm}), $d_{mM}^{\ell}(\theta_E)$ and $d_{MM'}^{\ell}(\theta)$
are related to the Wigner D-matrices by
\begin{equation}
D^{\ell}_{m'm}(\phi,\theta,\psi) = e^{-im'\phi}d^{\ell}_{m'm}(\theta)e^{-im\psi}.
\end{equation}

Analogously, the transpose convolution in harmonic space is given by
\begin{equation}
\label{eq:tconv_harm}
y^{P\ast}_{\ell m} = \sum_{m'm''} d^{\ell}_{mm'}(\theta_E)
d^{\ell}_{m'm''}(\theta) b_{\ell m''}^{P\ast} T_{mm'm''}
\end{equation}
where $P=I,G,C$.

\subsection{PReBeaM Implementation}

Now we outline the algorithmic steps taken to 
make a map from a TOD vector by PReBeaM. 

First we construct the 
right-hand side of equation (\ref{eq:norm_harm}) in two steps: converting TOD to a $T_{mm'm''}$ 
array and applying ${\mathrm{\bf A^T}}$.  
$T_{mm'm''}$ is constructed by transpose interpolating the TOD vector {\bf d}.  
The transpose
interpolation of the TOD vector onto the $T_{mm'm''}$ grid is akin to a binning step,
where each element of the TOD is mapped, via interpolation weights, to several 
elements of the $T_{mm'm''}$ cube according to the orientation and position of
that data point in the scanning-strategy.  The interpolation scheme is described in greater
detail in \S\ref{sec:interpol}.
Next, we transpose convolve the beam coefficients $b_{\ell m}$ with $T_{mm'm''}$ according to
equation (\ref{eq:tconv_harm}).

Once the right-hand side has been computed, we use the 
conjugate gradient iterative method to solve
equation (\ref{eq:normal}).  With each iteration, the coefficient matrix ${\mathrm{\bf A^TA}}$, is applied
using the following procedure:
\begin{enumerate}
\item Apply the convolution operator, {\bf A}, to project the sky $a_{\ell m}$ on to 
the convolution grid $T_{mm'm''}$
\item Inverse Fourier transform over the first two indices of $T_{mm'm''}$ to
get $T_{\Phi_2,\Theta,m''}$ (we omit the transform over $m''$ as it is 
incorporated in the interpolation scheme)
\item Forward interpolate from $T_{\Phi_2,\Theta,m''}$ to a TOD vector
\item Transpose interpolate from the TOD vector to a new ring set $T'_{\Phi_2,\Theta,m''}$
\item Fourier transform over the first two indices of $T'_{\Phi_2,\Theta,m''}$ to get
$T'_{mm'm''}$
\item Apply the transpose convolution operator, ${\mathrm{\bf A^T}}$, to project the ring set $T'_{mm'm''}$
back into a new sky $a_{\ell m}$ vector
\end{enumerate}

\subsection{Polynomial Interpolation and Zero-Padding}
\label{sec:interpol}
\begin{figure}[h]
\includegraphics[ width=.4\textwidth,keepaspectratio,
	angle=0,origin=1B]{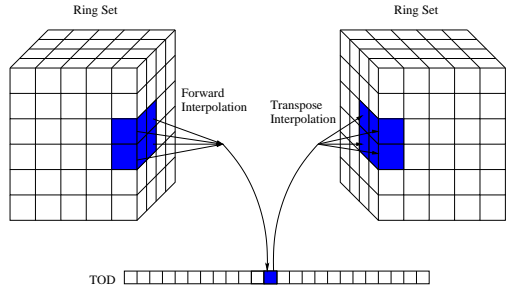}
\caption{Forward interpolation from ring set to TOD element and transpose interpolation
from TOD element to ring set.}
\label{fig:interpol}
\end{figure}

PReBeaM uses the same polynomial interpolation as implemented in the Level-S software 
used to generate the simulation TODs and as described in \citet{R06}.  The objective
of forward interpolation is to construct a TOD element at a particular co-latitude, longitude
and beam orientation using several elements of the ring set $T$ and their corresponding weights.  
Transpose interpolation operates in exactly the opposite manner as the forward interpolation:
distributing a single element in the TOD to multiple elements of the ring set. 
This is done using the same weights calculated for the forward interpolation.  The entire
operation of interpolation and transpose interpolation from ring set to TOD and back again
is depicted in figure \ref{fig:interpol}

PReBeaM also includes the option
to zero-pad during the FFT and inverse FFT steps.  This means that the working array (either 
$T_{mm'm''}$ or $T_{\Phi_2,\Theta,m''}$) is enlarged and padded with zeroes out 
to $\ell_{max,pad}>\ell_{max}$. This has the effect of decreasing 
the sampling interval.
We found that the combined effects of small-order polynomial interpolation (order 1 or 3) 
and zero-padding
of $2\times\ell_{max}$ or $4\times\ell_{max}$ 
dramatically reduced the residuals in our maps.

\subsection{Parallelization Description}
PReBeaM employs a hierarchical parallelization scheme using 
both shared-memory (OpenMP) and distributed-memory (MPI) types of
parallelization.  The map-making was performed on the NERSC computer Bassi.  Bassi processors
are distributed among compute nodes, with 8 processors per node.  OpenMP tasks occur within a node
and MPI tasks occur between nodes. 

\begin{figure}
\includegraphics[ width=.4\textwidth,keepaspectratio,
	angle=0,origin=1B]{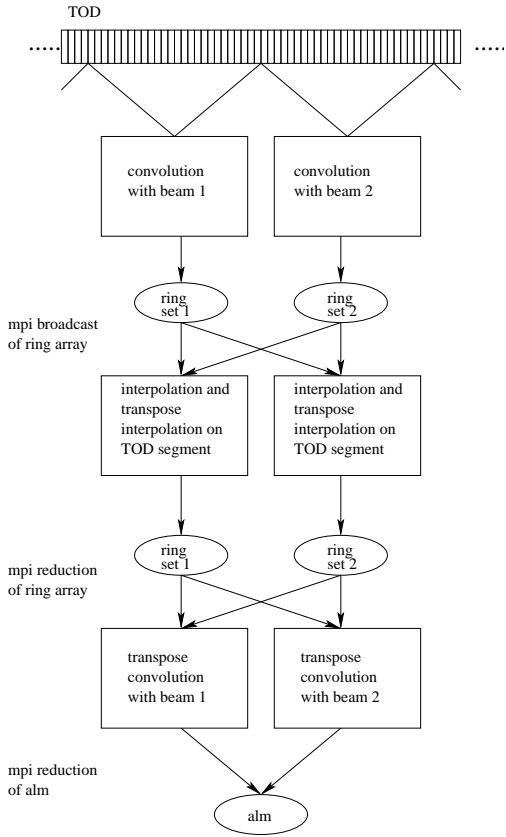}
\caption{Depiction of hybrid parallelization scheme used in PReBeaM.  Rectangles represent work 
done on a node, ellipses represent data products, and 
arrows represent transfer of data.  The work done within a node (convolution, interpolation and
their transpose operations) is parallelized using OpenMP.  
This shows a slice of two head nodes though the algorithm may operate on many more nodes.}
\label{fig:parallel}
\end{figure}

We show a diagram of our hybrid parallelization scheme in Figure \ref{fig:parallel}.
The full TOD and pointings are divided equally between the nodes for input and
storage of pointings.  
Within an iteration loop, four head nodes are designated to perform the convolutions, while the 
remaining active nodes are dedicated to the interpolation routines.  Each
 of these four nodes performs the convolution of the sky with one of the four detectors.
The resulting arrays are then distributed to all nodes for interpolation over the 
segment of data stored there
and then gathered
back onto the designated nodes for transpose convolution.  Finally, the $a_{\ell m}$ are 
summed, using MPI task mpi\_reduce, into a single $a_{\ell m}$ on a
single node; this is the new estimate for the sky vector.

This particular scheme was devised so that the four distinct convolutions that must occur
(one sky with four different beams) can take place simultaneously, while the pointings are 
distributed among as many nodes as possible for maximum speed in interpolation.  Both convolution
and interpolation and their transpose operations make use of all processors on a node by using OpenMP
directives.

\section{Simulations and Beams}
\label{sec:sims}
The simulated \textsc{Planck} data on which PReBeaM was run was generated by the 
\textsc{Planck} CTP 
working group for the study of the performance and accuracy of five
 map-making codes summarized in the Trieste paper \citep{Trieste}.  
\textsc{Planck} will spin at a rate of approximately one rpm, with an angle between the spin
 axis and the optical axis of $\sim 85^{\circ}$.  We used the cycloidal scan
 strategy in which the spin axis follows a circular path with a period of six months, 
and the angle between the spin axis and
 the anti-Sun direction is $7.5^{\circ}$.
TODs were generated for 366 days for the four 30 GHz Low Frenquency Instrument (LFI) 
detectors.  At a 
sampling frequency of 32.5 Hz, this corresponds to 1.028x$10^9$ samples per detector, 
for a total of over 65 Gb of data and pointings.  
The simulated data 
also included the effects of variable spin velocity and nutation (the option
 to include the effects of a finite sampling period was not included).

The data was simulated with elliptical having with a geometric mean FWHM of 
$32.^{\prime}1865$ and ellipticity (maximum FWHM divided by minimum FWHM) 
of 1.3562  and 1.3929 for each pair of horns. The widths and orientations 
of the beams were different; this was referred to as {\it beam mismatch} in the Trieste paper. 
In spherical harmonic space, the simulation beams were 
described up to a beam $m_{max}$ of 14.  The same beams were used in PReBeaM to solve 
for the map, although we allowed the beam asymmetry parameter $m_{max}$ to vary.

\section{Results and Discussion}
\label{sec:results}
For this paper, we make temperature and polarization maps from simulated 
one-year observations 
of the four 30 GHz detectors of the \textsc{Planck} LFI.  
We examine two cases: CMB signal only and CMB plus uncorrelated (white) noise.  Foreground signals and correlated
noise properties will be examined in a future paper. 
The 30 GHz data was an optimal choice for this analysis
because the low sampling rate and resolution minimize the data volume, while the large 
beam ellipticity allows us to demonstrate the full potential of our beam deconvolution technique.

PReBeaM operates entirely in harmonic space, solving for and producing as output $a_{\ell m}$.
For visualization purposes, maps were made from $a_{\ell m}$'s out to $\ell_{max}$ 512 
and at the Healpix \citep{healpix} resolution 
of nside 512 ($\sim 7^{\prime\prime}$ pixel size).  Most of the results presented in 
this paper were 
attained with an FFT zero-padding
of factor four, an interpolation order of 3, and an asymmetry parameter of $m_{max}=$4 (we note
where the parameters differ from this).
To compare with the input signal, 
a reference map representing the true sky was created by smoothing the 
input $a_{\ell m}$ by a Gaussian beam of FWHM = $32.^{\prime}1865$.  Similarly, our 
regularizer $G$ 
(in equation (\ref{eq:regular})) was set to have a FWHM of $32.^{\prime}1865$ to match 
this smoothing.
As noted in \S\ref{sec:sims}, the same data we use here has been processed by five map-making
codes in \cite{Trieste}.  We have chosen to compare our results with the analogous results
from Springtide, one of the codes in this study.  Springtide was chosen, out of the 
five codes, because it
is the map-making code installed in and used by the \textsc{Planck} Data Processing Centers 
for the HFI and LFI
instrument.  It is sufficient to compare with Springtide only as no significant differences in 
accuracy were found between codes (with similar baselines and in the absence of noise) 
\citep{Trieste}.
In the absence of noise, Springtide is algorithmically akin to a straight-forward binning of 
the TOD into a
sky pixel map.  Because we are using Springtide to represent all non-beam-deconvolution methods  
we will refer to the Springtide maps as the {\it binned} maps.

\begin{figure*}
\includegraphics[ width=.5\textwidth,keepaspectratio,
	angle=0]{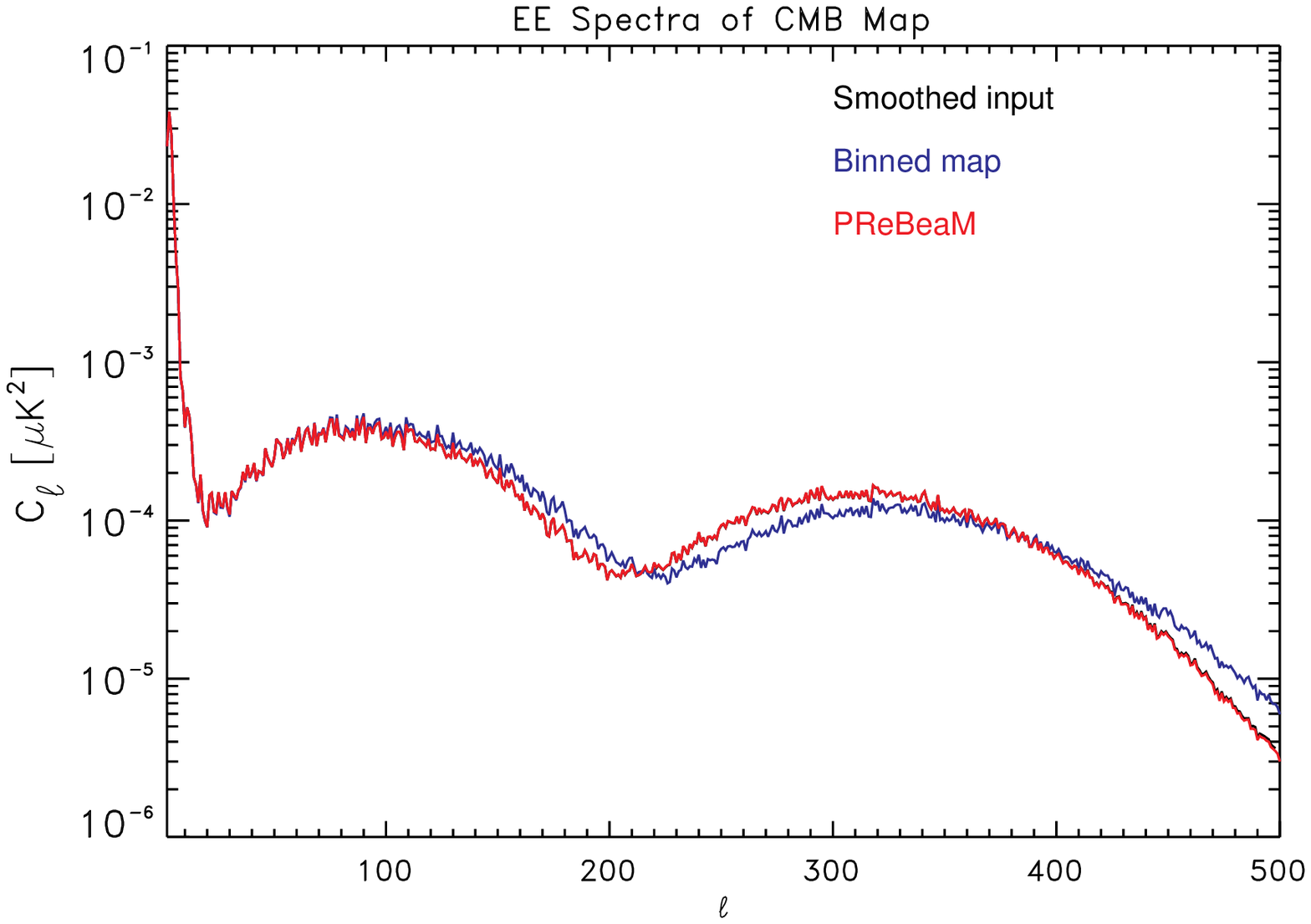}
\includegraphics[ width=.5\textwidth,keepaspectratio,
	angle=0]{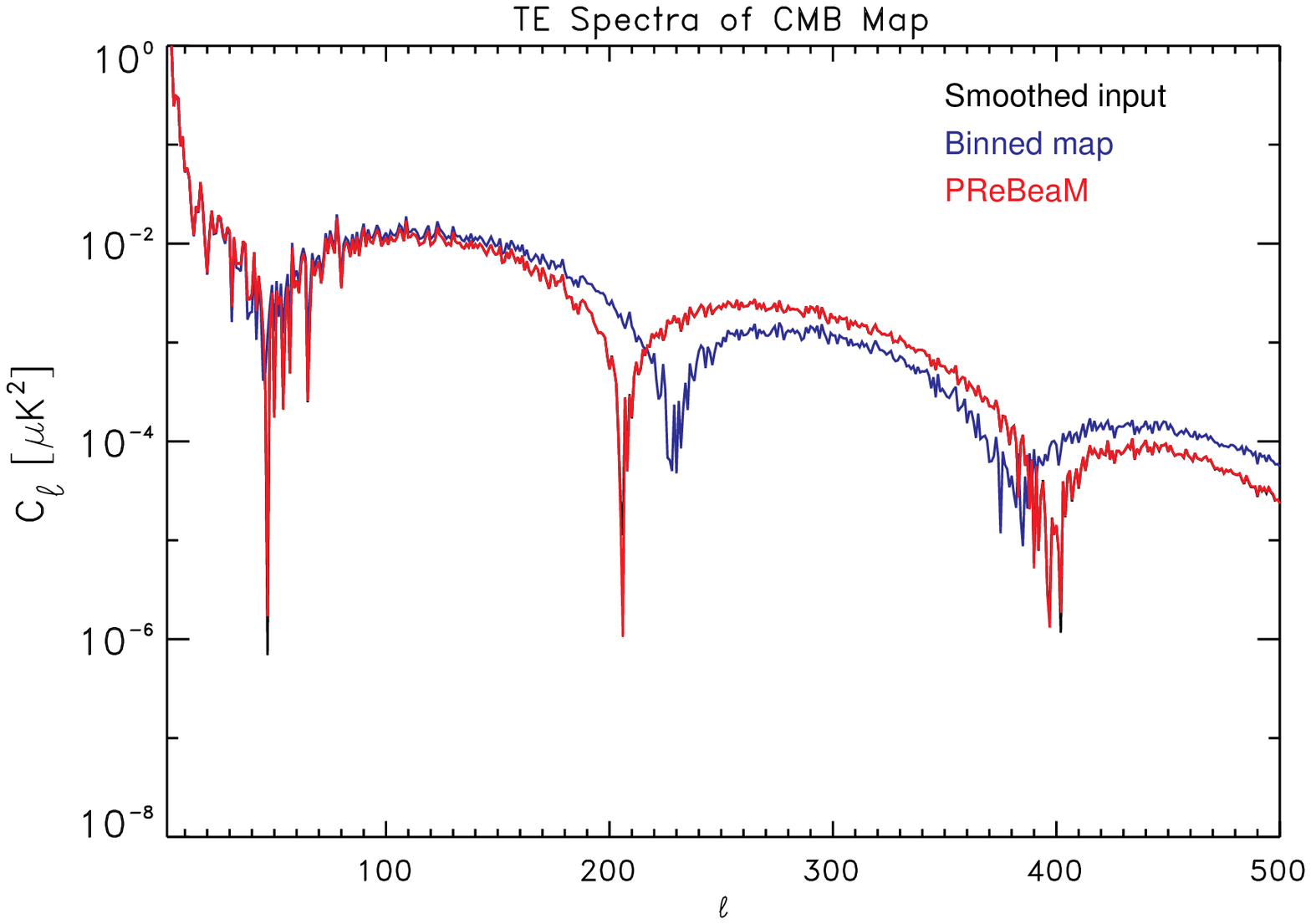}
\includegraphics[ width=.5\textwidth,keepaspectratio,
	angle=0]{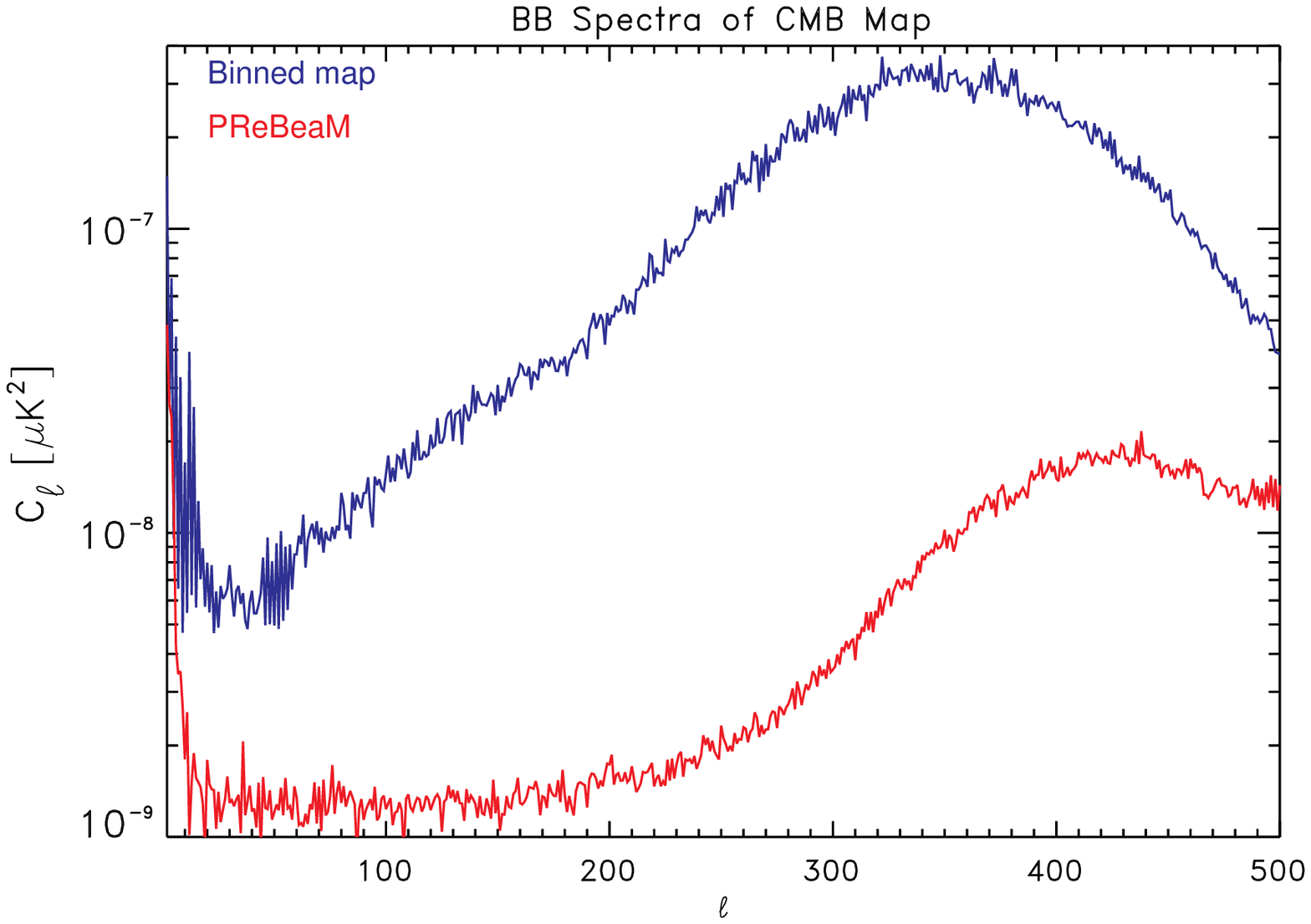}
\caption{EE, TE, and BB spectra of smoothed input map (black curve), binned map (blue curve) 
and PReBeaM (red curve).  The EE and TE spectra show
the effect of temperature-to-polarization cross-coupling seen in the binned map 
spectra as shifts in the peaks and valleys and absent from the PReBeaM spectra.  The input BB spectra 
is absent from the BB plot since the input B-modes were zero.
TT spectra are omitted since differences in the three spectra are not apparent in this representation.}
\label{fig:cls}
\end{figure*}

We begin by examining the spectra in the binned map, PReBeaM map and
the smoothed input map shown in Figure \ref{fig:cls}.  The 
effect of the beam mismatch
is clearly seen where the peaks and valleys of the binned map
spectra have been shifted towards higher multipoles.  
The detectors measure different Stokes I which translates
to artifacts in the polarization map.  Deconvolution suppresses leakage from temperature 
to polarization as evidenced by the PReBeaM spectra which overlays the input spectra.
This shift is expected
to remain apparent in the TE spectra of non-beam-deconvolved maps even in the 
presence of noise because of larger temperature
signal and the temperature-to-polarization cross-coupling.

\begin{figure*}
\includegraphics[ width=.5\textwidth,keepaspectratio,
	angle=0]{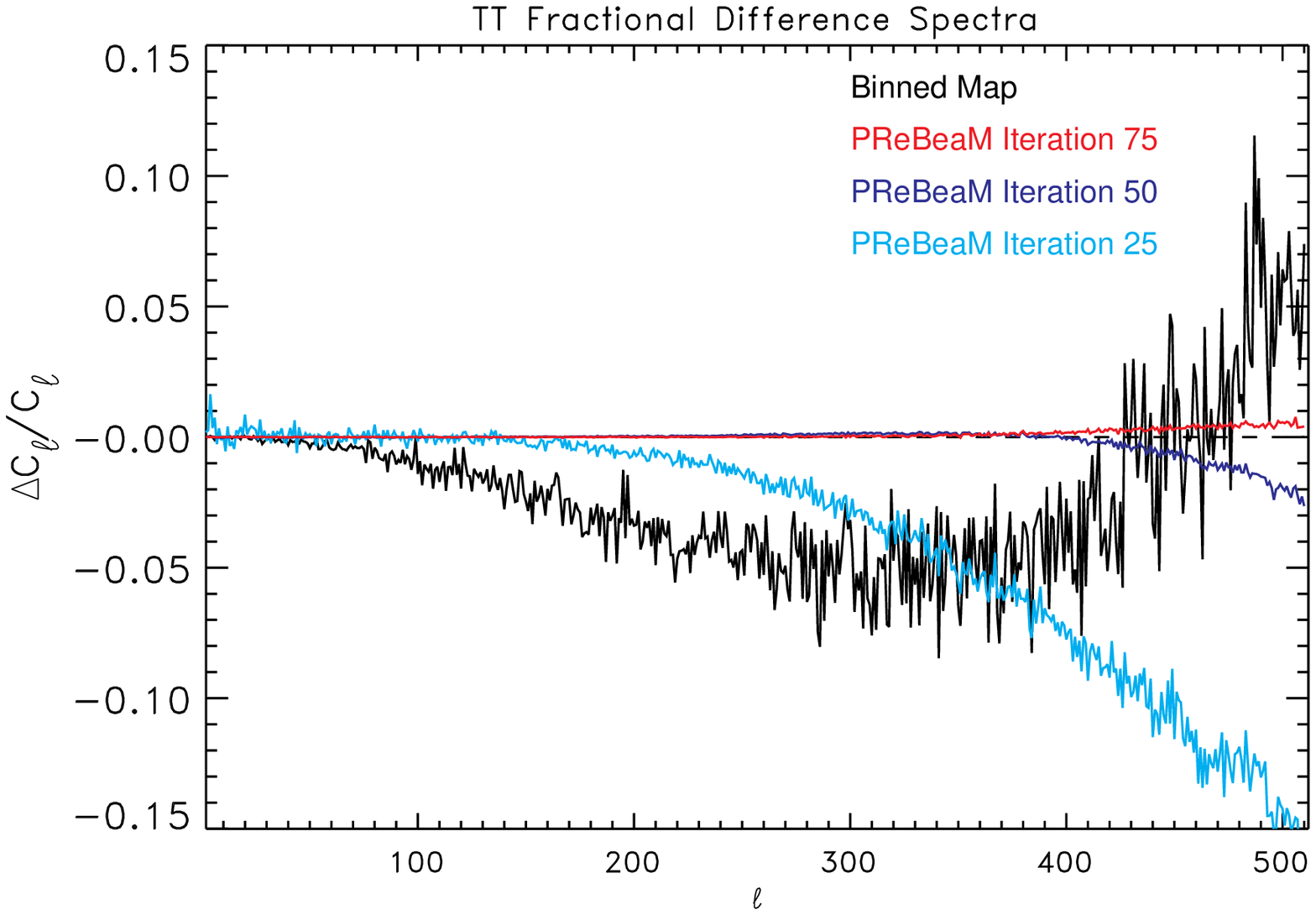}
\includegraphics[ width=.5\textwidth,keepaspectratio,
	angle=0]{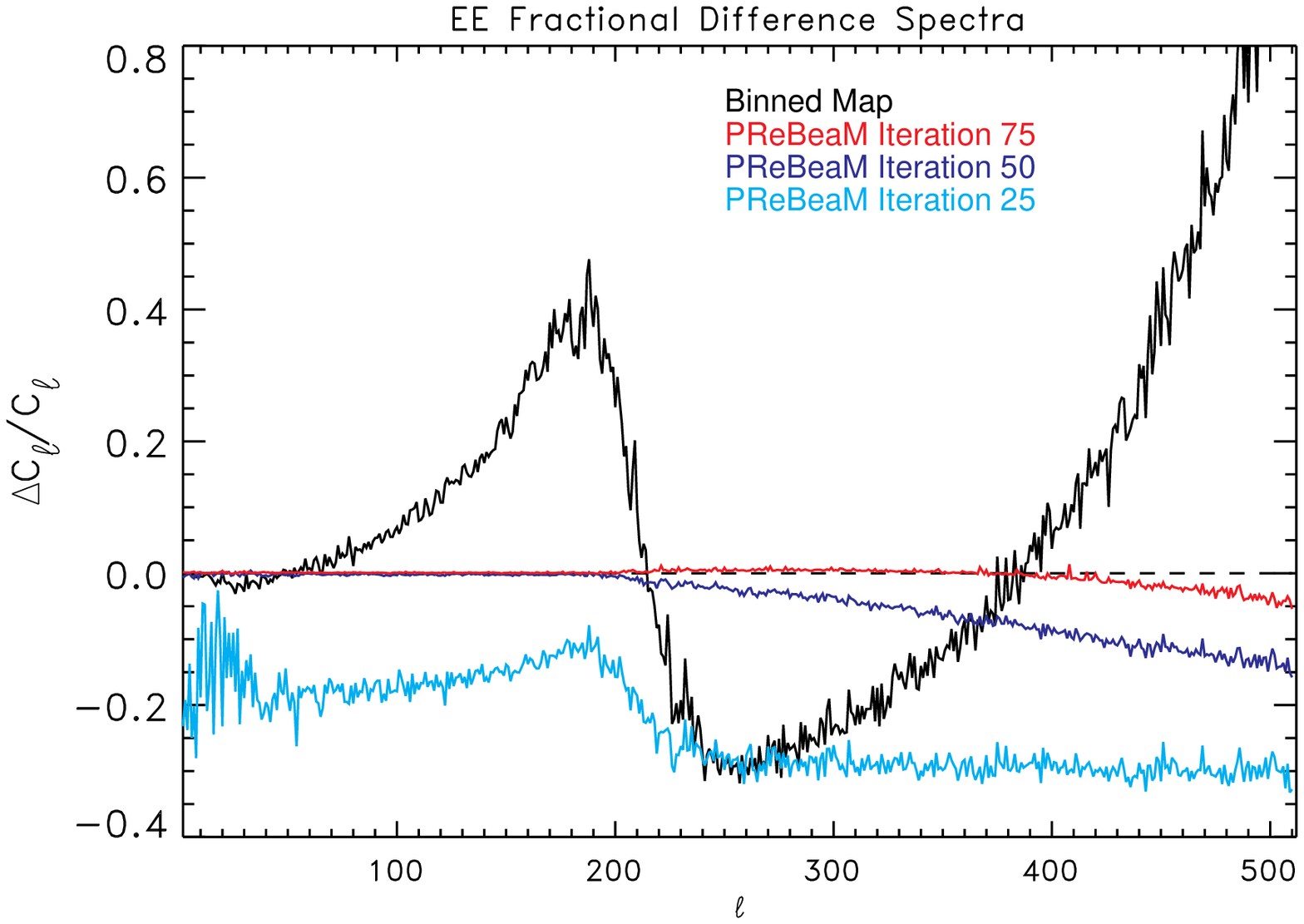}
\includegraphics[ width=.5\textwidth,keepaspectratio,
	angle=0]{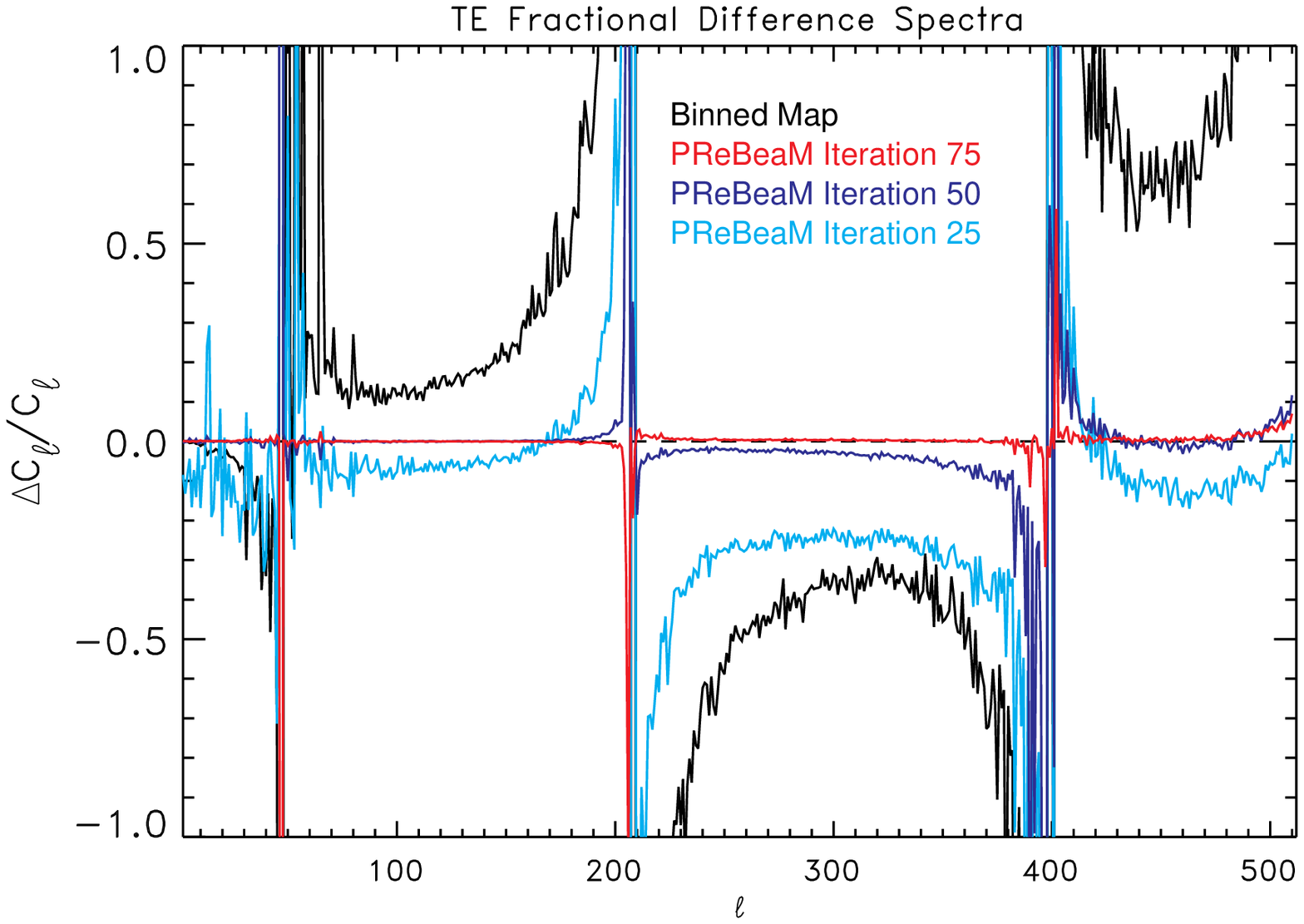}
\caption{Fractional difference in power spectrum for PReBeaM (red, blue, and cyan curves) and
the binned map (black curve) for TT, EE, and TE.  Spectra for PReBeaM are shown as a function
of number of iterations to demonstrate convergence.}
\label{fig:cls_frac}
\end{figure*}

The fractional difference in the angular power spectrum 
(defined as $(C_{\ell_{out}}-C_{\ell_{in}})/C_{\ell_{in}}$) of the input and output maps
is shown in Figure \ref{fig:cls_frac}.  We show the fractional difference spectra for the 
TT, EE, and cross-correlation TE signals, omitting the BB spectra since $C^{BB}_{\ell}$ 
is zero in the simulation of the CMB map .
 The results for PReBeaM are shown at three 
intervals: the 25th, 50th and 75th iterations.  This shows the behavior of the power
spectra as PReBeaM converges on the solution. 
The beam mismatch effect is also seen in Figure \ref{fig:cls_frac}, where the fractional difference
in the PReBeaM spectra lie closer to zero than the binned map spectra over the full range of multipole moments
for EE and TE. 

\begin{figure*}
\includegraphics[ width=.5\textwidth,keepaspectratio,
	angle=0]{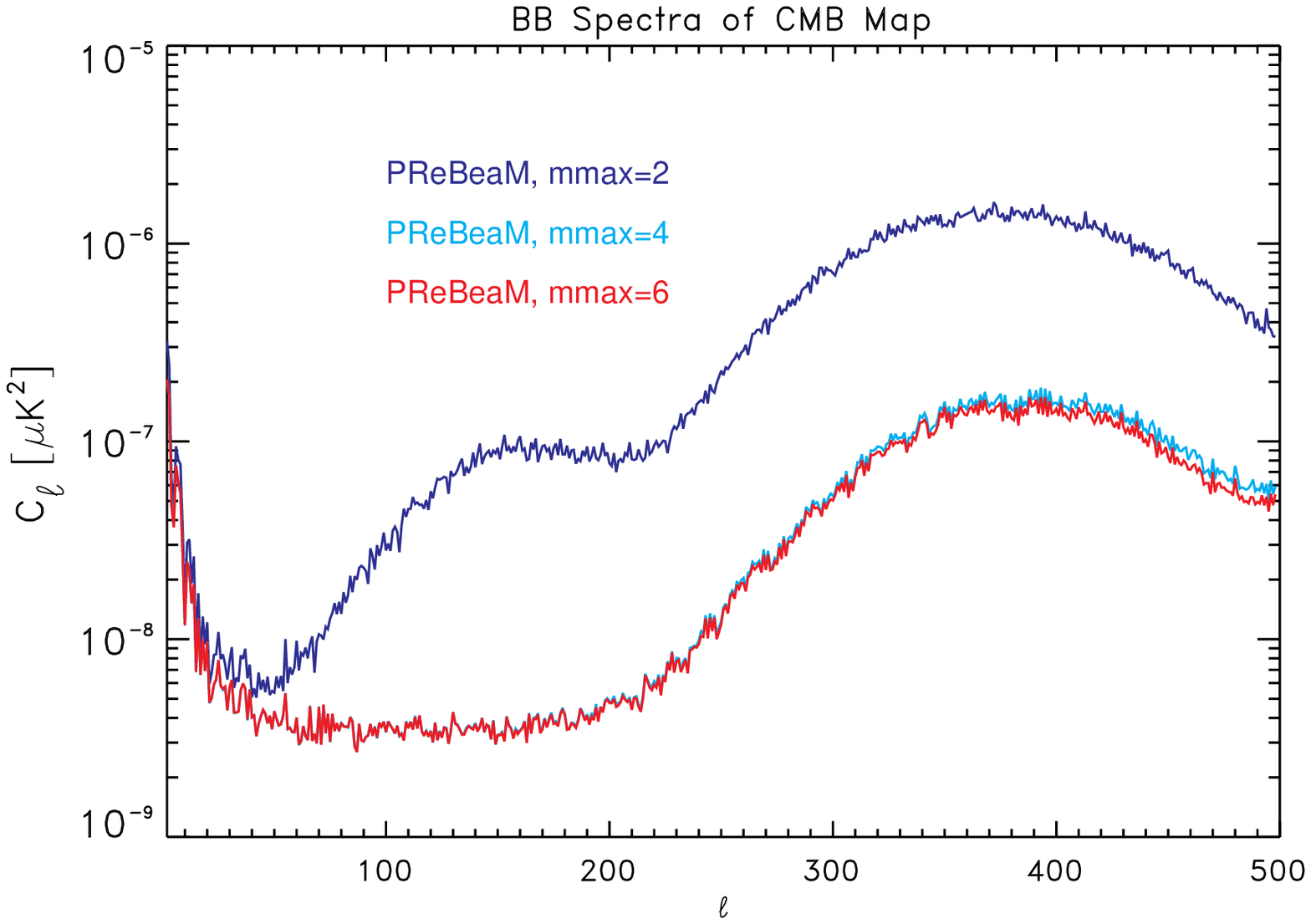}
\caption{BB power spectra as a function of asymmetry parameter $m_{max}$ for $m_{max}=$ 2 (blue curve)
, 4 (cyan curve), and 6 (red curve).  The input BB spectra was zero so
the smallest output BB spectra is most desirable.  In this run, the PReBeaM input parameters interpolation order 
and zero-padding were set to 1 and 2, respectively.}
\label{fig:mmax}
\end{figure*}

As described earlier, PReBeaM allows for variation in the asymmetry parameter $m_{max}$.  We examined 
the performance of PReBeaM as a function of $m_{max}$, setting it to 2, 4 and 6.  A remarkable improvement
in the power spectra was found by increasing $m_{max}$ from 2 to 4, while an increase from 4 to 6 only resulted
in marginal improvements.  This effect is best seen in the BB power spectra as shown in Figure \ref{fig:mmax}.
Thus, while the input TOD was simulated with a beam having an $m_{max}$ cut-off of 14, PReBeaM 
operates optimally at an $m_{max}$ of just 4, thereby allowing us to capitalize
 on the computational property that PReBeaM scales as $m_{max}$.

\begin{figure*}
\includegraphics[ width=.5\textwidth,keepaspectratio,
	angle=0]{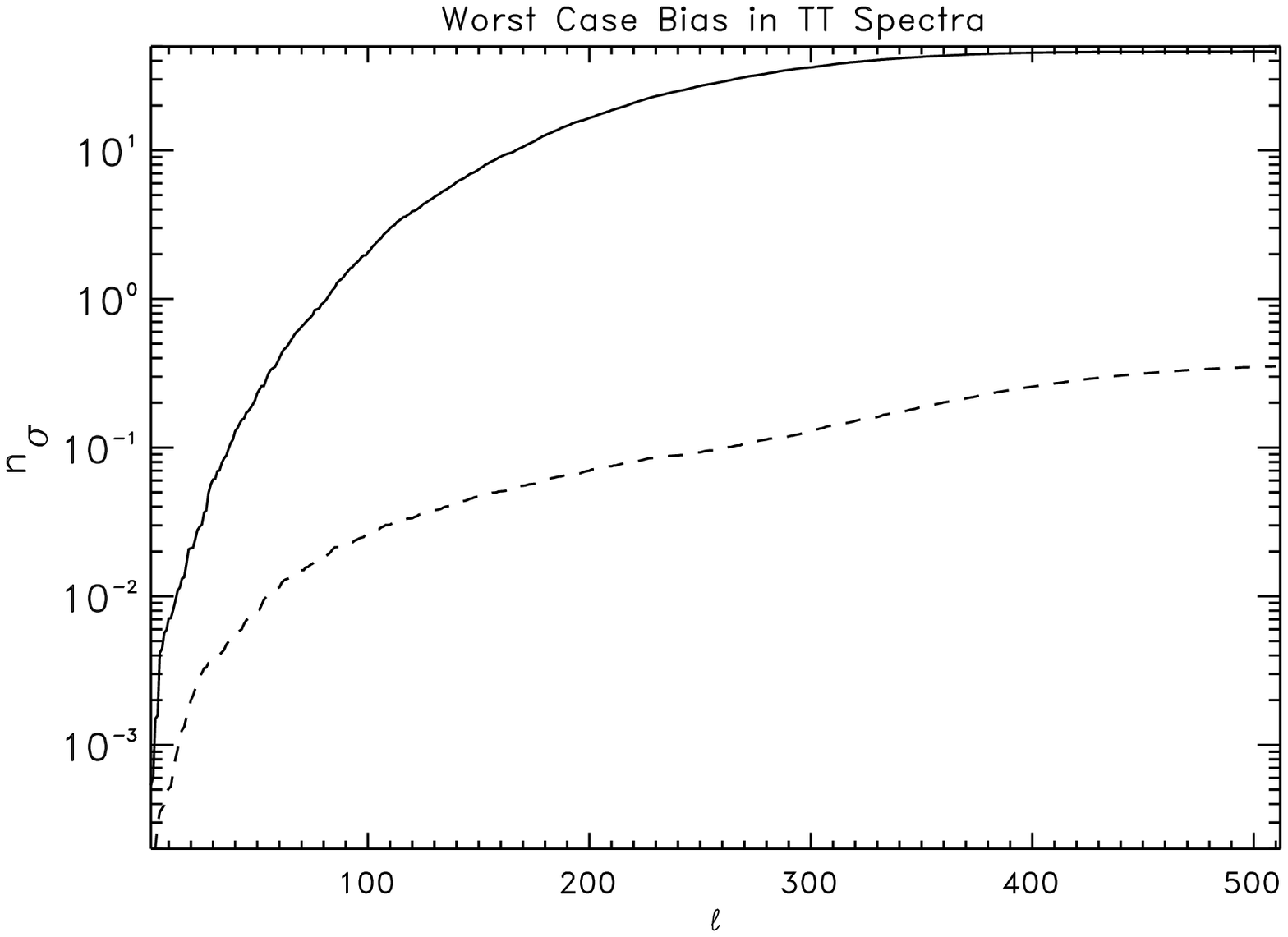}
\includegraphics[ width=.5\textwidth,keepaspectratio,
	angle=0]{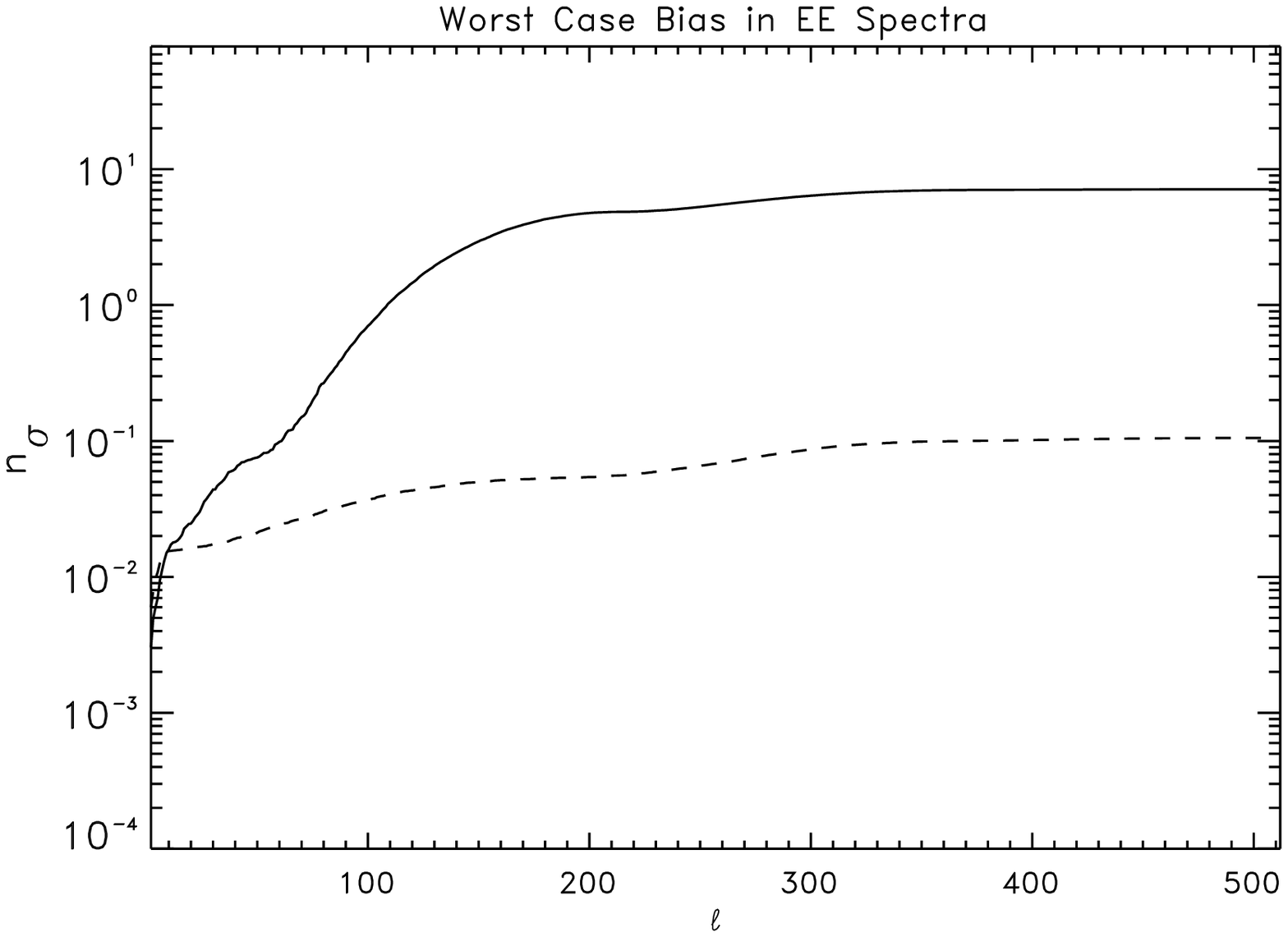}
\includegraphics[ width=.5\textwidth,keepaspectratio,
	angle=0]{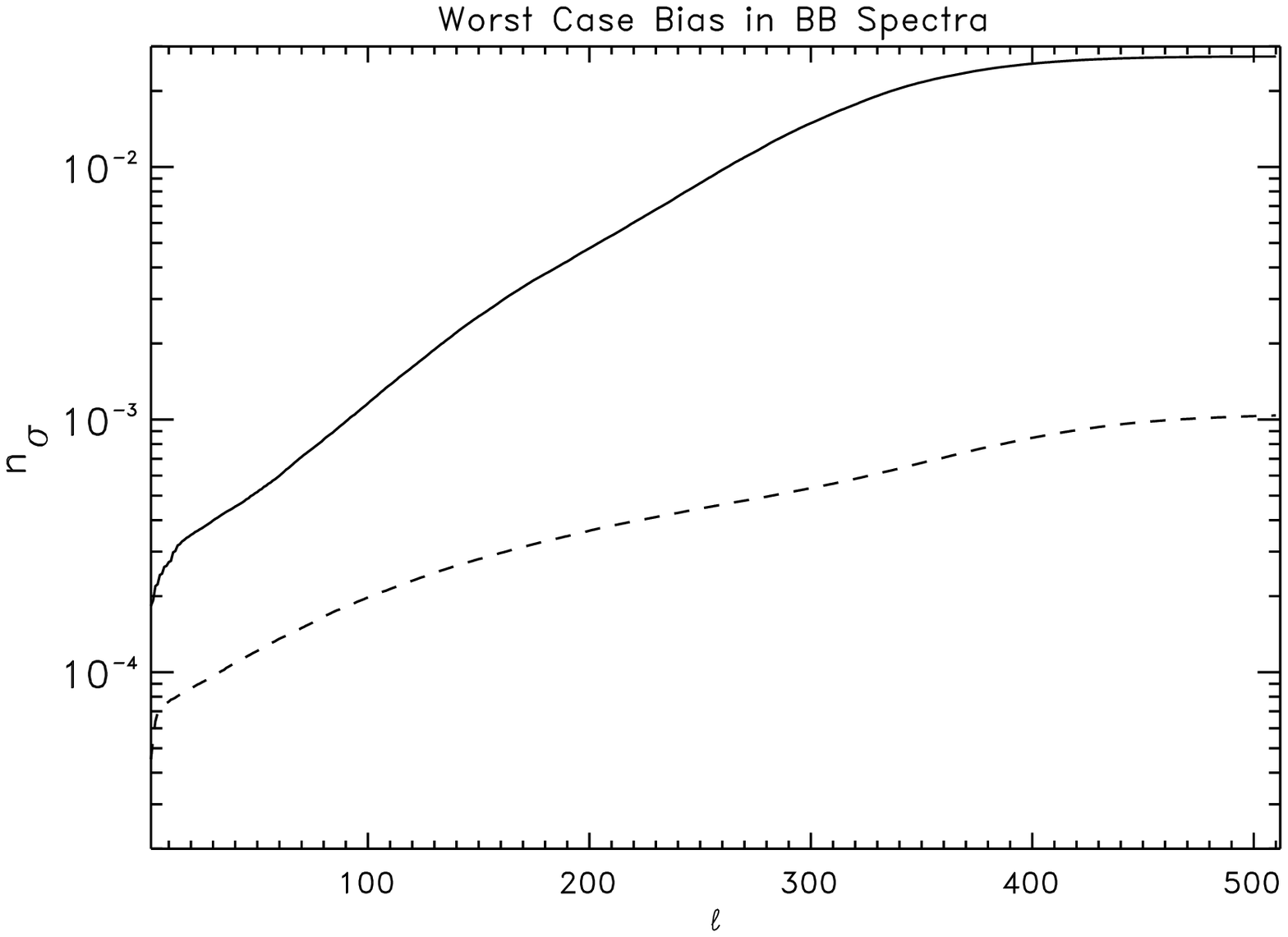}
\includegraphics[ width=.5\textwidth,keepaspectratio,
	angle=0]{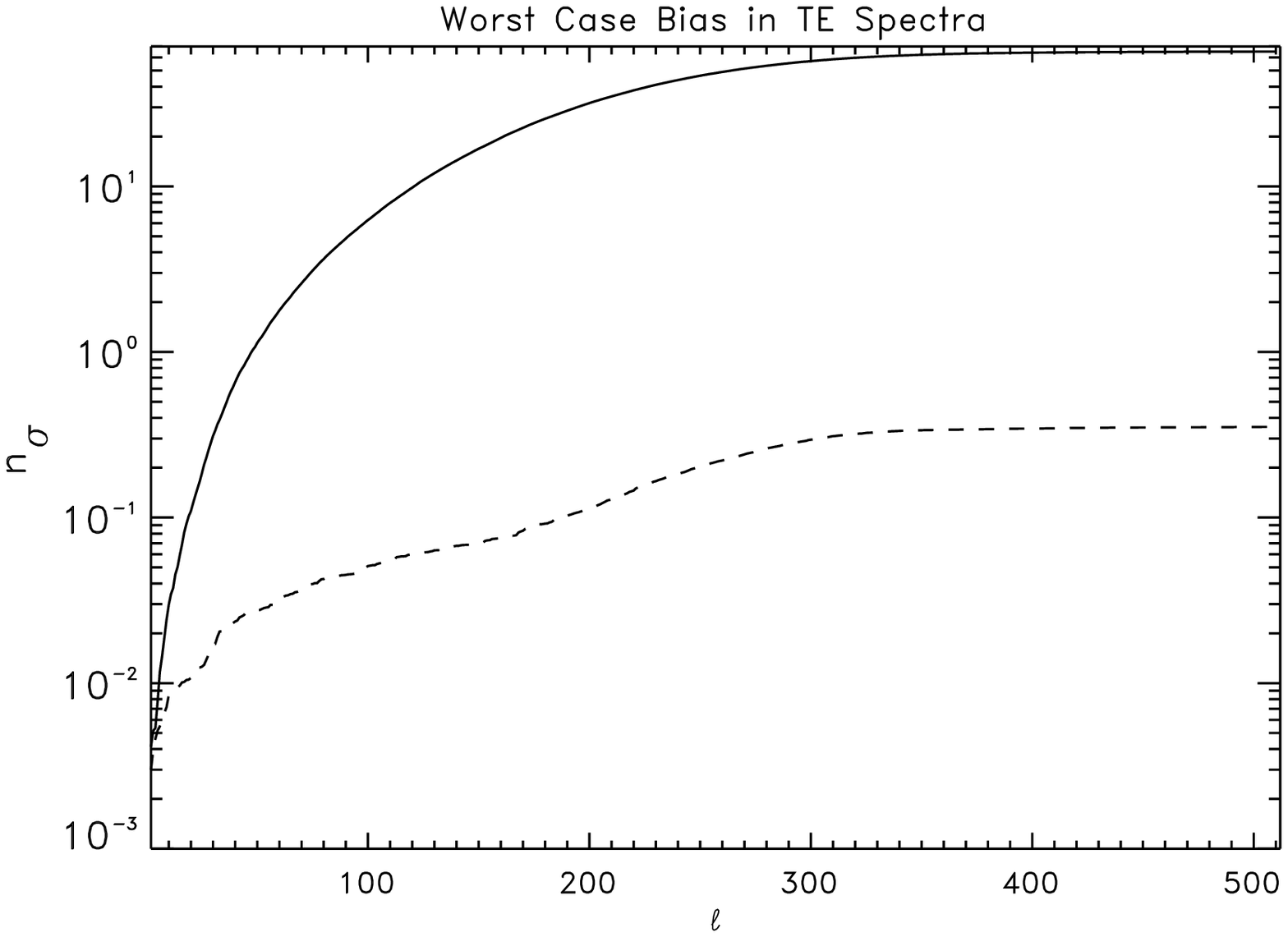}
\caption{Worst case bias in estimation of cosmological parameters due to 
errors in the power spectra of PReBeaM (dashed curve) and due to the errors in the power
spectra of the binned map (solid curve).}
\label{fig:nsigs}
\end{figure*}

We define a quantity called $n_{\sigma}$ 
\begin{equation}
n_{\sigma_{\ell}} = \sum_{\ell^{\prime}=2}^{\ell}\frac{|\Delta C_{\ell^{\prime}}|} {\sigma_{Planck_{\ell^{\prime}}}}
\end{equation}
which we use to quantify the maximum, or worst-case bias beam systematics could induce 
in a cosmological parameter that happened to be degenerate with that parameter.
The quantity $\sigma_{Planck}$ is the expected one-sigma errors for the LFI 30 GHz channel,
computed as the diagonal elements of the covariance matrix for the 
simulated input spectra, assuming a sky fraction of 0.65.
The $n_{sigma}$ values are plotted in Figure \ref{fig:nsigs} and show that 
PReBeaM reduces the worst case bias due to untreated beam systematics from tens of sigma to 
much less than one sigma over the entire $\ell$ range.

\begin{figure*}
\includegraphics[ scale=.3, keepaspectratio,
  angle=90]{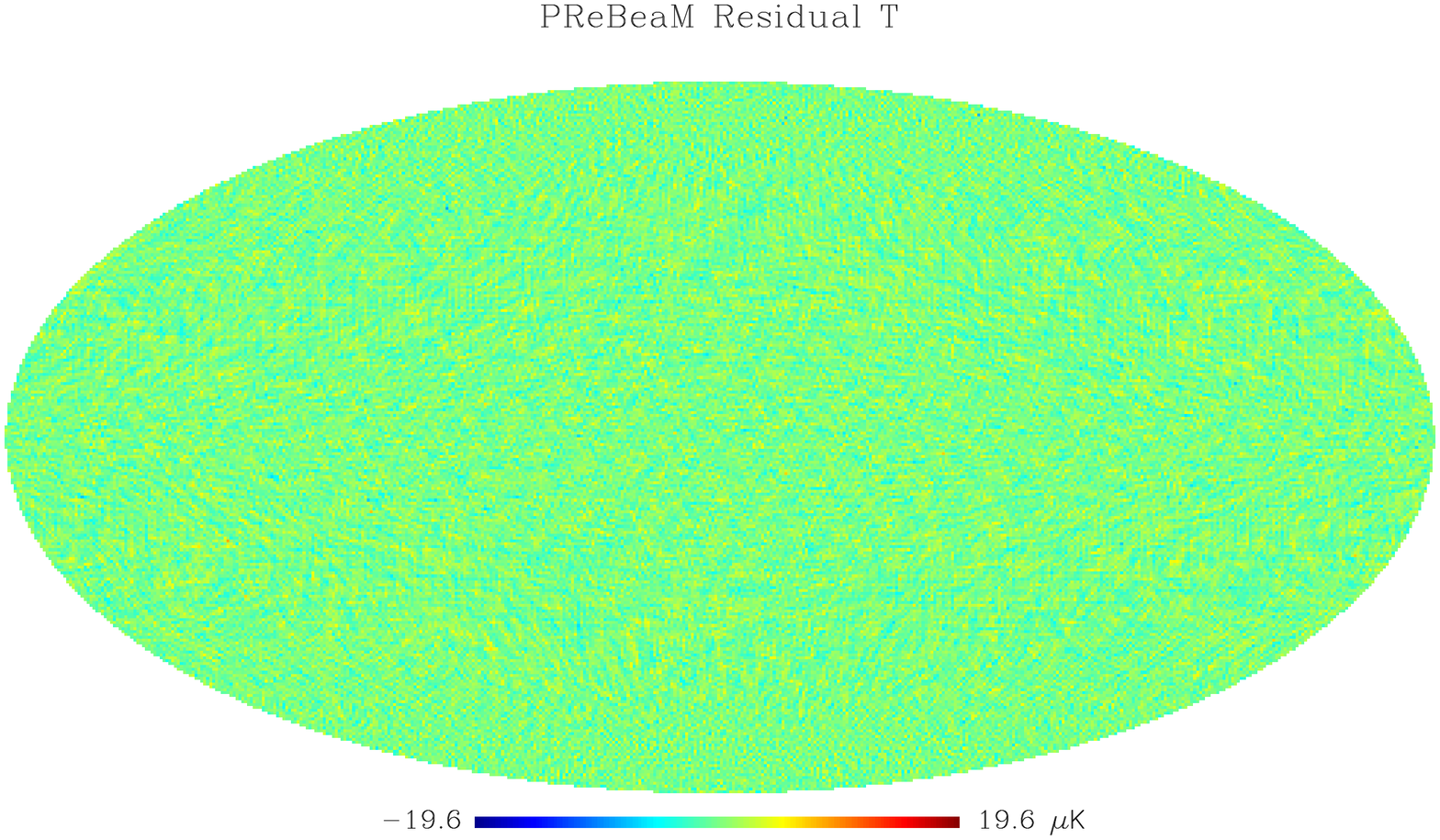}
  \includegraphics[ scale=.3, keepaspectratio,
  angle=90]{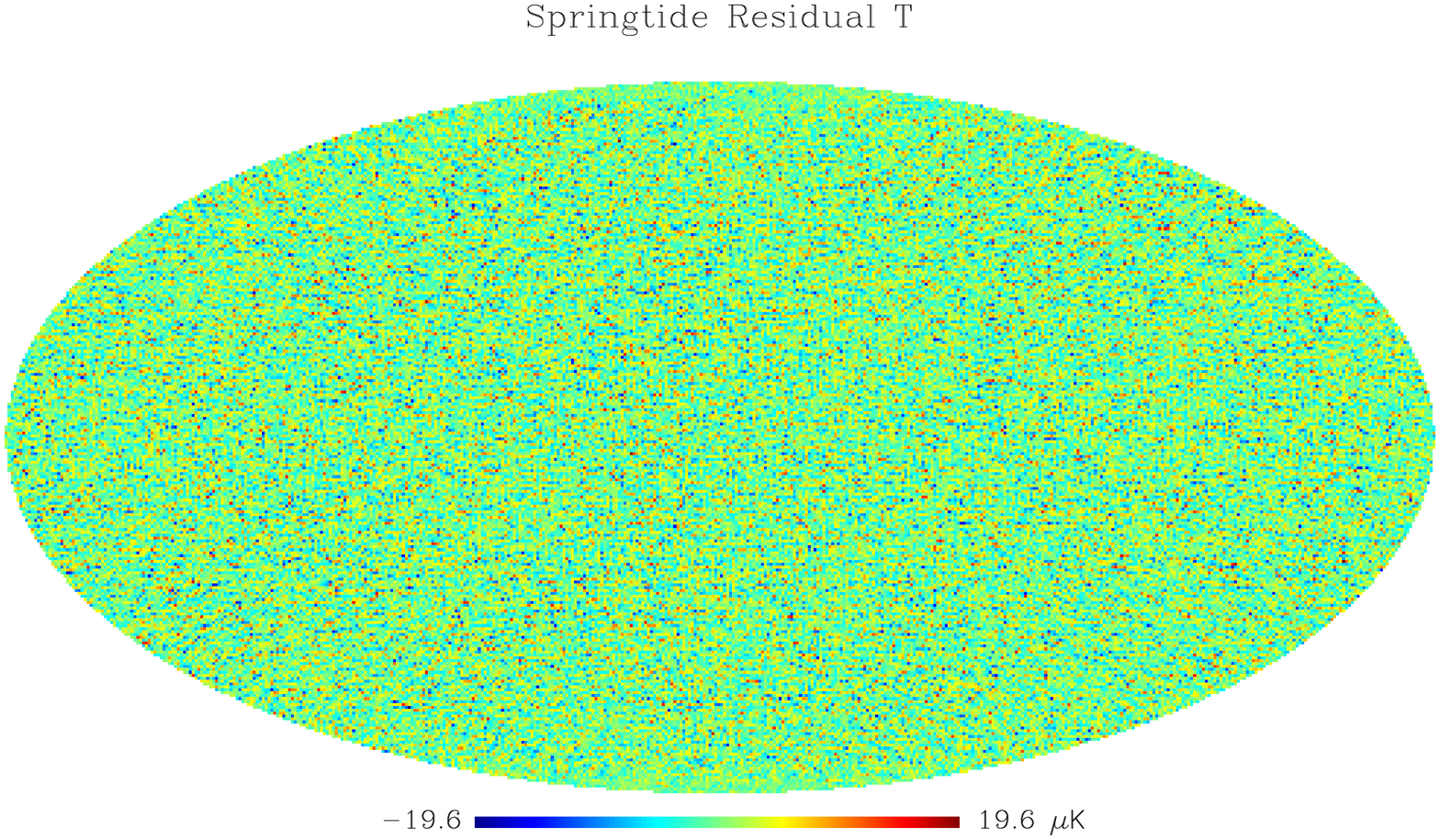}
  \includegraphics[ scale=.3, keepaspectratio,
  angle=90]{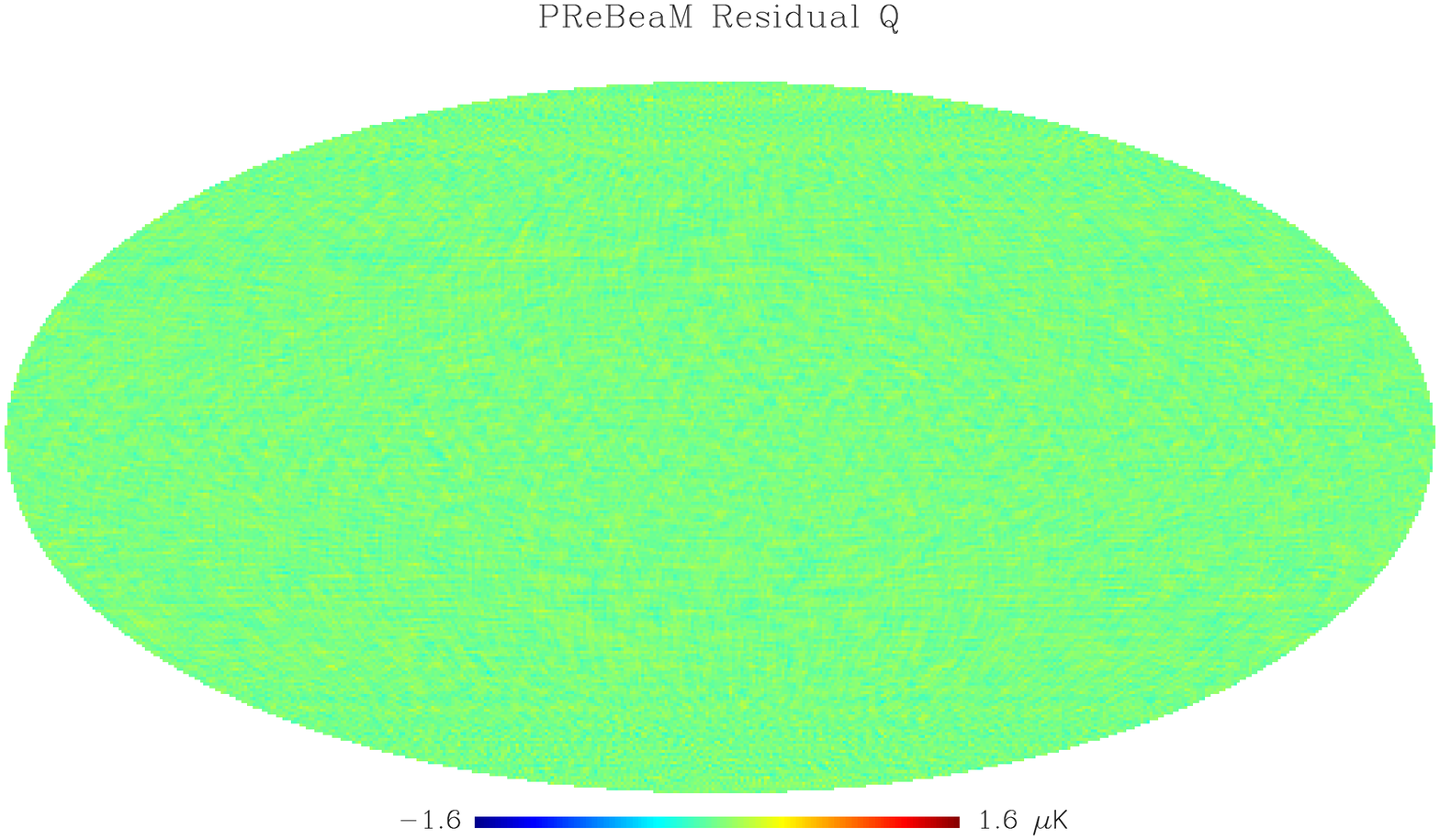}
  \includegraphics[ scale=.3, keepaspectratio,
  angle=90]{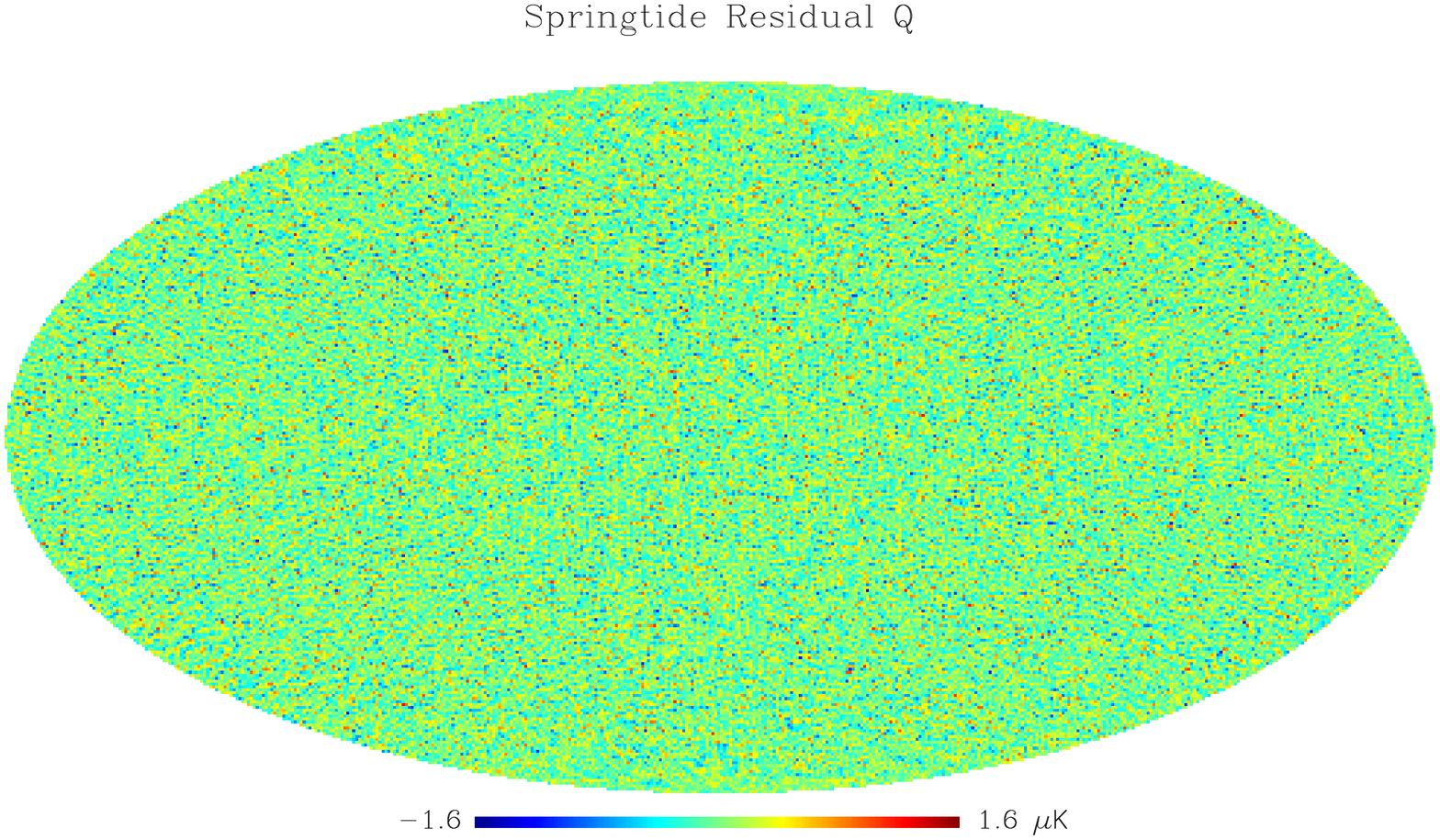}
  \includegraphics[ scale=.3, keepaspectratio,
  angle=90]{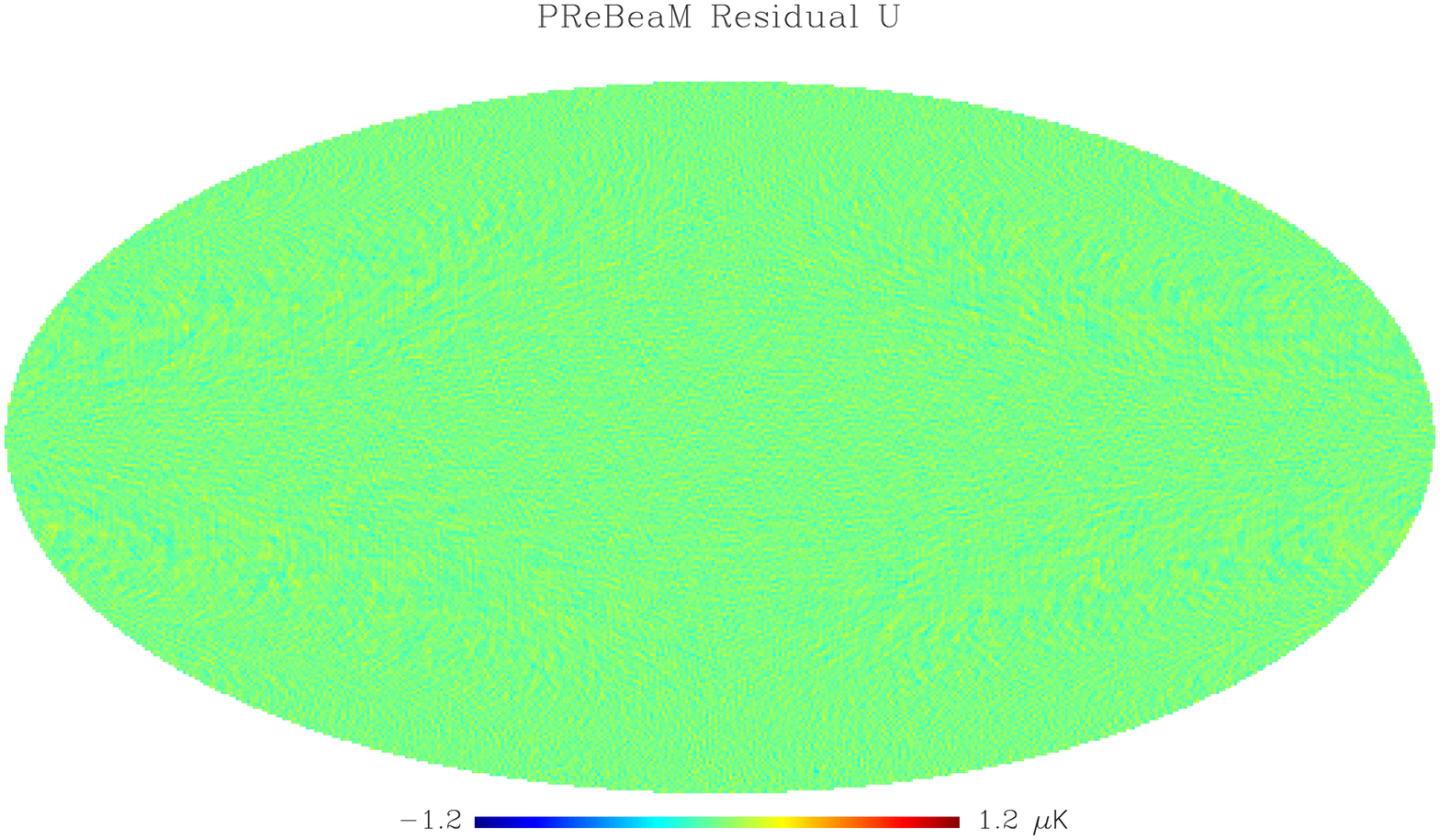}
  \includegraphics[ scale=.3, keepaspectratio,
  angle=90]{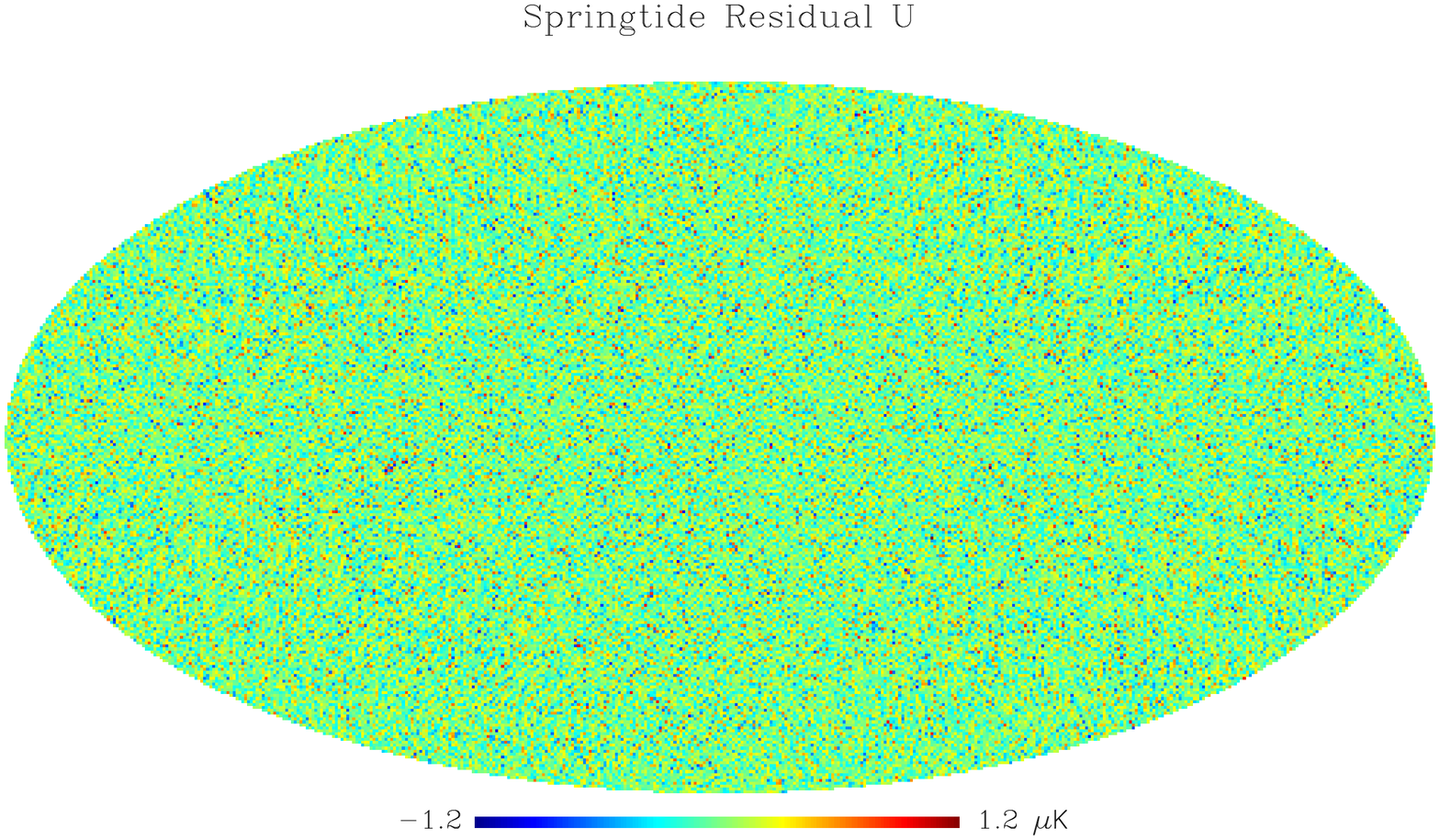}
\caption{The residuals between the input reference sky and PReBeaM output (left column) and the
residual between the input reference sky and the binned map (right column) for Temperature (T), 
and the Stokes Q, and U parameters.}
\label{fig:resmaps}
\end{figure*}

We examine the resulting temperature and polarization (Q and U) maps.
The output map for both PReBeaM and the binned map was 
subtracted from the smoothed input map at the same resolution to make the 
residual maps shown in Figure \ref{fig:resmaps}.  PReBeaM residuals were plotted
on the same color scale as the binned map, showing that PReBeaM attained smaller
residuals for both temperature and polarization.

As a final test, we run PReBeaM on TOD containing CMB signal and white noise and compare
with the smoothed input CMB spectrum and the analogous results from Springtide (in this
case we refer to Springtide directly since this is not simply a binned map).
The level of the uncorrelated noise is specifed in the detector database and has 
a nominal standard deviation per sample time of $\sigma=1350\mu K$ \citep{Trieste}.
PReBeaM achieves a noticeably superior fit to the input spectrum compared with Springtide
from $\ell\sim150$ to $\sim250$.  Assessing the relative performance of PReBeaM and 
Springtide in more detail would require performing Monte Carlo averages. We focus on the 
TE spectrum since the improvement is visible even for a single simulation. For the other 
spectra PReBeaM performs as least as well as Springtide but the detailed difference are more 
difficult to assess without a Monte Carlo study.

\begin{figure*}
\includegraphics[ width=.5\textwidth,keepaspectratio,
	angle=0]{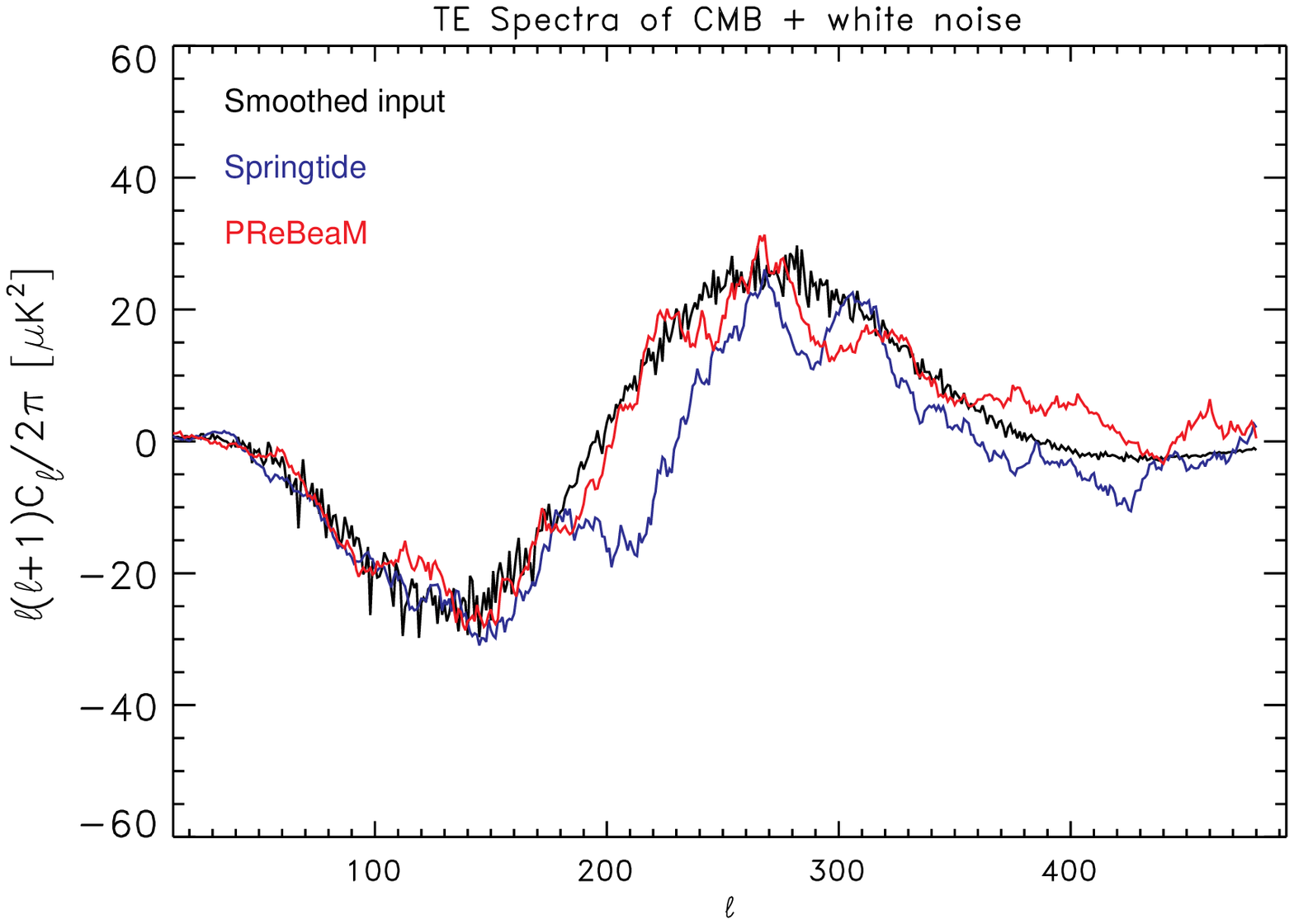}
\caption{TE spectrum of CMB and white noise for Springtide (blue curve) and PReBeaM (red curve).
The smoothed input map is shown in black.  
Following the example in \cite{Trieste}, 
we reduce $\ell$ to $\ell$ variation by filtering the spectra by a sliding average 
($\Delta\ell=20$).  
In this run, the PReBeaM input parameters interpolation order 
and zero-padding were set to 1 and 4, respectively. While PReBeaM performs at least as well as Springtide in the TT, EE, and BB spectra, we omit 
these spectra since the detailed differences are difficult to assess without an in-depth 
Monte Carlo study.}
\label{fig:whitenoise}
\end{figure*}

\subsection{Computational Considerations}
The computational costs and advantages of our method should be noted.  To perform a 
convolution up to $\ell_{max}$ requires $\mathcal{O}(\ell^3_{max}m_{max})$ for the 
general case.  Since $m_{max}$ is bounded by $\ell_{max}$, the cost never 
scales worse than $\mathcal{O}(\ell^4_{max})$ and is only $\mathcal{O}(\ell^3_{max})$ for 
the symmetric beam case.  By comparison, a
brute force computation in pixel space would require $\mathcal{O}(\ell^5_{max})$.  In this study,
data was simulated with beams having an asymmetry parameter of $m_{max}=14$, but maps were made using
a cut-off value of $m_{max}=4$ in PReBeaM.  We have demonstrated that computational 
cost can be conserved while still achieving the benefits of beam deconvolution

It was found that an increase in the zero-padding factor from two to four produced 
superior results over
an increase in the interpolation order from one to three.
An optimal run of PReBeaM will therefore include the largest zero-padding possible given
machine memory constraints in conjunction with a polynomial interpolation of order one or three.
This is advantageous since the time spent in an FFT is nearly negligible and affected
only minimally with an increase in zero-padding.  In contrast, time for interpolation scales
as interpolation-order-squared and as this is a TOD-handling step, it dominates over any cost
incurred by convolutions.  In the case of the results shown here, interpolation steps consume
 more than 90\% of the wall-clock time per iteration.

The results produced here were generated using 12 nodes on NERSC computer Bassi 
(making use of all 8 processors
per node) and was complete in about 29 wall-clock hours, for a total of 2797-CPU hours.  
The maximum task memory was 20 GB on a single node.  

\section{Conclusion}
\label{sec:conclusion}

We have found that PReBeaM has outperformed the standard binned noiseless map
 using two measures: spectra and 
residual maps.  We examined the fractional differences in the spectra
and found markedly smaller differences in the PReBeaM spectra versus the binned map spectra across
a range of multipole moments.  We find that 
map-making codes which do not deconvolve
beam asymmetries lead to significant systematics in the polarization power spectra 
measurements.  The temperature-to-polarization cross-coupling due to beam asymmetries 
is manifested as shifts in the peaks and valleys of the spectra.  These shifts are absent from the
PReBeaM spectra.  We translated the errors found in the power spectra to an estimate of the 
statistical significance of the errors in a parameter estimation resulting from these spectra, 
which we call $n_{\sigma}$.  This analysis showed that the worst case parameter bias due to 
beam-induced power spectrum systematics could be tens of sigma while PReBeaM reduces the 
risk of parameter bias due to beam systematics to much less than 1 sigma
We also found the I, Q, and U component residual maps to be smaller for PReBeaM than 
for the binned map, implying smaller map-making errors.

We have presented here the first results from PReBeaM for a straightforward test cases of 
CMB only and CMB plus white noise,
and including only the effects of beams in the main focal plane.  However, there is great
potential for using PReBeaM to remove or assess systematics due to the combination of foregrounds
and beam side lobes.  Systematics introduced by side lobes will appear on the largest scales, 
potentially impeding the detection of primordial B-modes on the scales where they are most 
likely to be measured.
We have already shown for temperature measurements \citep{AW04} that our
deconvolution technique can be used to remove effects due to side lobes.  Future work will 
examine the noise properties of PReBeaM maps and will include foregrounds from extragalactic 
sources and diffuse Galactic emission.

\end{document}